\documentclass[10pt,a4paper]{article}

\usepackage{times}
\usepackage{amssymb,amsmath,xcolor}

\usepackage{epsfig,wrapfig}
\usepackage{epstopdf} % for pdflatex

\usepackage{fancyhdr}

\usepackage[lmargin=2.5cm, rmargin=2.5cm,tmargin=3.50cm,bmargin=3.50cm]{geometry}
\usepackage{indentfirst}

\usepackage{titlesec}
\titleformat{\section}[hang]
  {\centering}{\thesection}{1ex}{\normalsize \textsc}%%
\titleformat{\subsection}[hang]
  {}{\thesubsection}{1ex}{\normalsize \textit}%%

\renewcommand{\thesection}{ \normalsize \textnormal{\Roman{section}.}}
\renewcommand{\thesubsection}{\normalsize \textnormal{\textsc{\textit{\Alph{subsection}.}}}}

\def\e{\begin{equation}}
\def\f{\end{equation}}
\def\_#1{{\bf #1}}

\def\.{\cdot}

\def\Re{{\rm Re\mit}}

\begin{document}

%%% Title of paper
\title{\large \textbf{Large-Period Multichannel Metagratings For Broad-Angle Absorption}}
%
%%% Author(s) and affiliation
\def\affil#1{\begin{itemize} \item[] #1 \end{itemize}}
\author{\normalsize \bfseries Y. Yashno and \underline{A. Epstein}}
% you will of course remove the next line...
%\\ \small (List of authors - in 10 pt Times New Roman. Underline the presenter. Use a second line if necessary)}
%
\date{}
\maketitle
\thispagestyle{fancy} % header also to the first page
\vspace{-6ex}
\affil{\begin{center}\normalsize Technion - Israel Institute of Technology,\\
Andrew and Erna Viterbi Faculty of Electrical and Computer Engineering, Haifa 3200003, Israel\\
epsteina@ee.technion.ac.il
\end{center}}

%%% Abstract
\begin{abstract}
\noindent \normalsize
\textbf{\textit{Abstract} \ \ -- \ \
%%% Start here with text of abstract
We present an alternative scheme for obtaining effective power dissipation in planar composites, extending the recently proposed concept of metagrating (MGs), \emph{sparse} arrangements of polarizable particles (meta-atoms), to realize multifunctional absorbers. In contrast to typical metasurface solutions, where periodicities are limited to half of a wavelength at most to avoid high-order Floquet-Bloch modes, we purposely consider large-period MGs, relying on their proven ability to effectively mitigate spurious scattering. The absorption process is thus implemented via precise engineering of the mutual coupling between numerous individual scatterers fitting in the enlarged period, with these additional degrees of freedom further utilized to enforce the perfect absorption conditions for multiple excitation angles simultaneously. The resultant devices, utilizing a standard printed circuit board configuration obtained semianalytically while featuring relaxed fabrication demands, exhibit high absorption across a wide angular range, useful for radar cross section reduction and energy harvesting applications.}
\end{abstract}

\section{Introduction}
\label{sec:intro}
Promoting absorption of electromagnetic waves in planar structures has been a topic of active research for many decades, %\cite{Dallenbach1938, Salisbury1952}, 
due to its central role in a wide variety of applications, from solar cells and THz sensors to radar cross section reduction and wireless power transfer. %\cite{Knott2004, Min2010}. 
In the classical Dallenbach absorbers and Salisbury screens, perfect absorption is obtained in stratified media formations which exhibit impedance matching to free space \cite{Radi2015}. Realizing these via natural materials, however, would typically require a specific (resonant) thickness ($\approx\lambda/4$), and may result in limited bandwidth and acceptance angle. New avenues to realize low-profile absorbers have become possible with the emergence of metasurfaces (MSs), thin slabs composed of closely-packed subwavelength polarizable particles (meta-atoms). Treating the densely arranged elements as homogenized generalized sheet transition conditions (GSTCs) \cite{Tretyakov2003}, %\cite{Tretyakov2003,Kuester2003}, 
capacitive sheets (e.g., patch arrays) with prescribed surface impedances could be conveniently devised, enabling the needed impedance matching even with ultrathin grounded dielectric substrates \cite{Luukkonen2009}.

Nevertheless, typical absorption applications require wide-angle response to maximize the overall scattering suppression or energy harvesting potential. While the GSTCs may yield the abstract surface impedance values required to realize absorption for a given (single) excitation, they do not guarantee 
%do not indicate which physical meta-atom configurations would exhibit the required response 
the performance in the entire desired angular range. %\cite{Epstein2016_2}. 
Consequently, extended angular response is obtained via sophisticated meta-atom geometries, devised using heuristic intuitive approaches combined with full-wave optimization \cite{Begaud2018}. %\cite{Yahiaoui2017, Begaud2018}.
%to extend the absorber angular response, heuristic intuitive approaches combined with full-wave optimization are typically utilized to devise sophisticated geometries to replace the simple patches \cite{Yahiaoui2017, Begaud2018}. 
Alternatively, improvement can be achieved by harnessing specialized (e.g., spatially dispersive) meta-atoms, but these are not always straightforward to translate into practical designs \cite{Luukkonen2009, DelRisco2021}. %\cite{Tretyakov2003_1, Luukkonen2009, DelRisco2021}.
Importantly, all these homogenization-based MS solutions utilize closely-packed finely-discretized unit cells \cite{Tretyakov2003}, %\cite{Epstein2016_2}, 
posing fabrication challenges and restricting possible geometries. Additional degrees of freedom may be enabled by introducing spatial modulation to induce auxiliary surface waves as design parameters \cite{Epstein2016_3}, but supercell dimensions are typically limited to half of a wavelength to avoid emergence of propagating higher-order Floquet-Bloch (FB) modes \cite{Wang2020}. Thus, design freedom is still quite limited, requiring implementation via dense, deep-subwavelength, meta-atoms with high-resolution features.

Recently, an alternative paradigm for molding fields has been proposed, based on metagratings (MGs) \cite{Radi2022}. In contrast to MSs, these periodic devices feature sparse meta-atom distributions, not governed by homogenization. To synthesize MGs, reliable models to predict scattering from given scatterer arrangements, properly accounting for mutual coupling, are devised within the FB framework, yielding direct relations between the MG structure and the coupling to the various FB modes \cite{Tretyakov2003}. %\cite{Tretyakov2003, Radi2017}. 
Subsequently, beam manipulation is obtained by judicious engineering of the meta-atom constellation and geometry such that the resultant interference pattern matches the desired functionality. At microwaves, we have derived theoretically and verified experimentally such a semianalytical synthesis scheme to enable highly-efficient diffraction engineering using printed-circuit-board (PCB) MGs based on loaded-wire meta-atoms, yielding fabrication-ready designs without resorting to full-wave optimization \cite{Rabinovich2020}. %\cite{Epstein2017, Rabinovich2020}.

Herein, we extend the MG design formalism to enable realization of broad-angle planar absorbers, retaining its appealing semianalytical nature and relaxed realization constraints. To this end, the model is augmented to allow incorporation of lossy loads, and the desired MG layout is obtained by demanding that scattered power to all propagating (reflected) FB modes would be minimal. Importantly, as previously demonstrated, the meticulous control on the coupling to a large number of FB modes achievable by multilayer multielement MGs allows us to alleviate the conventional MS limitation on the periodicity, enabling introduction of additional meta-atoms in the augmented period $\Lambda$; these, in turn, provide 
%facilitating utilization of large spacings between neighbouring elements. Beyond the manufacturing-related benefits, such configurations 
%enable introduction of additional meta-atoms (additional degrees of freedom) to the period, providing 
means to tailor the MG response for multiple excitation angles \emph{simultaneously} following a multichannel scattering approach \cite{Wang2020}. %\cite{Asadchy2017_1, Wang2020}.
%These results, verified via simulations in commercial solvers, establish an alternative paradigm for highly-effective absorption, relying on orchestrated scattering from multiple particles in large-period sparse composites. 
Beyond the simpler structure and reduced design time,
%clear advantages of the reduced complexity and semianalytical approach,
%relaxed structural complexity and reduced %(full-wave-optimization free) 
%design time,
these results establish a fundamentally different paradigm for broad-angle absorption, relying on orchestrated scattering from multiple particles and delicate interference engineering in large-period sparse composites. %to obtain highly-effective multifunctional performance.

\section{Theory, Results, and Discussion}

\begin{figure}[t!]
\centering \includegraphics[width=15cm]{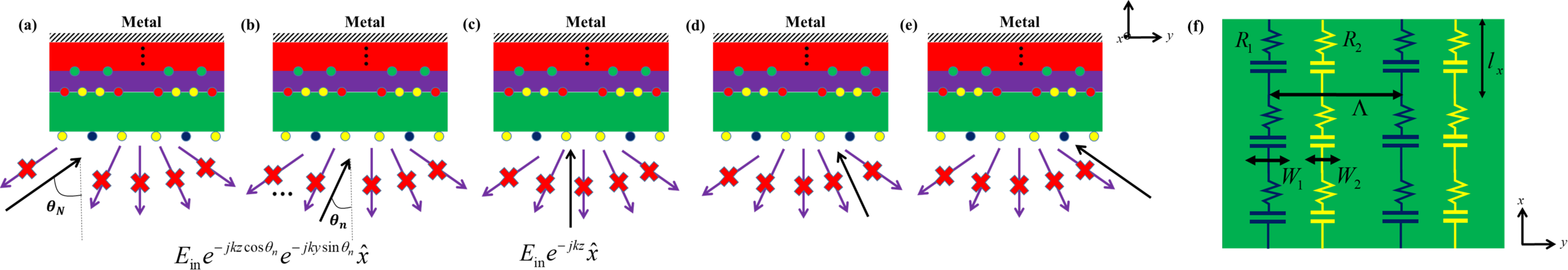}
\caption{Physical configuration of a multilayer multielement MG absorber \cite{Rabinovich2020}. (a)-(e) Symmetric multichannel scenario with $2N+1$ different input angles $\theta_n=\arcsin(n\lambda/\Lambda)$. (f) Top view of a single layer with two loaded-wire meta-atoms per period, featuring printed capacitors of width $W_k$ and lumped resistors $R_k$.} \label{fig:physical_configuration}
\end{figure}

We consider a $\Lambda$-periodic MG configuration excited by transverse electric ($E_z=0$) fields (Fig. \ref{fig:physical_configuration}). The MG features $K$ wires per period, positioned at $(y,z)=(d_k,h_k)$ along the interfaces of a $M$-layer conductor-backed PCB \cite{Rabinovich2020}, each loaded by printed capacitors of width $W_k$ and lumped resistors of resistance $R_k$ repeating every $l_x\!\!\ll\!\!\lambda$ along $x$. To facilitate the desired multichannel response, we devise a symmetric MG configuration realizing simultaneous perfect absorption for $2N+1$ excitation angles $\theta_n\!=\!\arcsin(n\lambda/\Lambda)$, $n\!=\!-N,...,N$. 
Given that $N\lambda\!<\!\Lambda\!<\!(N+1)\lambda$, the FB theorem implies that for each of these angles of incidence, power can only be scattered to the other channels $\theta_n$ \cite{Wang2020}. 
Treating the loaded wires as induced current sources, the scattered fields for given $\theta_n$, $(d_k,h_k)$, $W_k$, and $R_k$, can be evaluated semianalytically by invoking Poisson's formula and Ohm's law \cite{Tretyakov2003, Rabinovich2020}. These relations can then be used to formulate a set of constraints requiring all scattering coefficients to vanish (perfect absorption) for all considered excitation angles, from which the optimal MG configuration is deduced. Considering symmetry and reciprocity, the number of overall constraints is greatly reduced from the general $(2N+1)^2$ to $(N+1)^2$, achieving the multifunctional response with a highly sparse configuration.
%\textcolor{red}{For each of these excitation angles, the model in \cite{Rabinovich2020} can then be used to evaluate the coupling to the various scattered modes as a function of the meta-atom position and loading, and the requirement for vanishing scattering coefficients (perfect absorption) can be formulated. Allowing also resistive load impedances, we utilize this model in a nonlinear equation solver (MATLAB \texttt{lsqnonlin}) to resolve the detailed MG configuration to enable effective scattering suppression for all considered excitations. Considering symmetry and reciprocity, the number of overall constraints required to guarantee perfect absorption for all $2N+1$ input channels is greatly reduced from the general $(2N+1)^2$ to $(N+1)^2$, enabling achieving the desired multifunctional response with a highly sparse configuration.}

\begin{figure}[t!]
\centering \includegraphics[width=16.0cm]{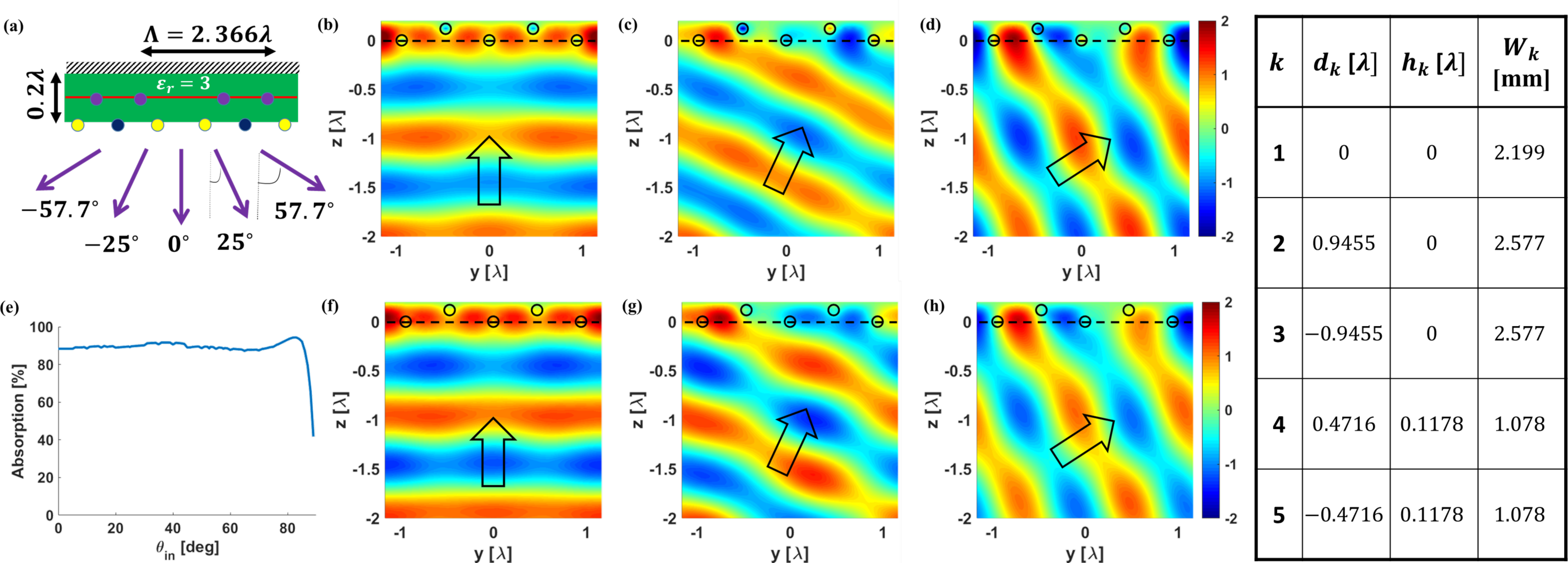}
\caption{Broad-angle MG absorber with N=2 (Fig. \ref{fig:physical_configuration}). (a) MG configuration as output by the semianalytical procedure (\textit{cf.} table). (b)-(d) Analytically predicted \emph{total} field distribution $\Re\{E_x(y,z)\}$ (one period) for the designated incidence angles (shown only for $\theta_n\geq 0$), (b) $\theta_\mathrm{in}=0^{\circ}$, (c) $\theta_\mathrm{in}=25^{\circ}$, and (d) $\theta_\mathrm{in}=57.7^{\circ}$. Meta-atoms are denoted by black circles; substrate/air interface is marked by a dashed line. (f)-(h) Corresponding total fields as recorded in full-wave simulation. (e) Full-wave simulated absorption as a function of the incident angle.} \label{fig:absorption_results}
\end{figure}

To demonstrate our methodology, we follow it to design the ($2N\!+\!1\!=\!5$) multichannel MG absorber of Fig. \ref{fig:absorption_results}(a), intended to absorb $20$ GHz plane waves incident from $\theta_{\pm 2}=\pm 57.7^{\circ}$, $\theta_{\pm 1}=\pm 25^{\circ}$, and $\theta_{0}=0^{\circ}$. Correspondingly, the MG periodicity is $\Lambda=\lambda/\sin 25^{\circ}$; it was found that $K=5$ meta-atoms per period are sufficient in this case. The nine unique constraints on the scattering matrix required to guarantee simultaneous perfect absorption from all input channels were resolved via MATLAB (\texttt{lsqnonlin}), yielding the optimal MG specifications. For lumped resistors of $R_k=37.7\Omega$, this semianalytical scheme placed the meta-atoms in two layers below the conductor at $z=0.2\lambda$, with wire coordinates and capacitor widths as listed in the table of Fig. \ref{fig:absorption_results} (symmetrical about $y=0$).
%$(d_1, h_1)=(0,0)$, $(d_2, h_2)=(-d_3,h_3)=(0.9455\lambda,0)$, $(d_4, h_4)=(-d_5,h_5)=(0.4716\lambda, 0.1178\lambda)$ and capacitor widths $W_1=2.199$ mm, $W_2=W_3=2.577$ mm, $W_4=W_5=1.078$ mm.  

Without any further optimization, we have defined this structure in CST, and simulated the MG response to each of the considered input channels. The results indicate that the designed MG succeeds in suppressing almost all back-reflected fields for each of the considered excitations, with absorption of $90\%$ ($97\%$), $89\%$ ($98\%$), and $88\%$ ($89\%$) obtained in full-wave simulations (analytical predictions) for $\theta_\mathrm{in}=\theta_{0}$, $\theta_\mathrm{in}=\theta_{\pm 1}$, and $\theta_\mathrm{in}=\theta_{\pm 2}$, respectively. Indeed, slightly increased specular reflection reduces power dissipation by up to $9\%$ compared to theory; however, as shown by the MG angular response extracted from CST for the entire range [Fig. \ref{fig:absorption_results}(e)], effective broad-angle absorption is obtained overall (above $88\%$ , $|\theta_\mathrm{in}|<85^{\circ}$). The excellent agreement between the theoretical [Fig. \ref{fig:absorption_results}(b)-(d)] and simulated [Fig. \ref{fig:absorption_results}(f)-(h)] fields further verifies the high fidelity of the analytical model, implying that larger periods can be considered to improve the angular coverage in the future.
%. These results indicate the great potential of the proposed concept, utilizing a practical sparse configuration in a large macro period while exhibiting, without any full-wave optimization, effective multi-angle absorption.

\section{Conclusion}
\label{sec:conclusion}
We have presented an alternative route for obtaining effective broad-angle absorption from low-profile structures, relying on \emph{large-period} PCB MGs comprised of \emph{sparsely distributed} loaded wires. By tailoring the mutual coupling between the scatterers via a detailed analytical model to suppress coupling to propagating FB modes, high absorption is maintained while providing space for numerous meta-atoms per period. 
These additional degrees of freedom are harnessed to achieve absorption from \emph{multiple} angles of incidence simultaneously, leveraging a symmetric multichannel approach for increased sparsity. Such insightful designs could lead to improved low-cost absorbers for radar evasion, fitting well also energy harvesting applications (power absorbed in a few loads).

\end{document}